\documentstyle[12pt]{article}
\begin{document}

\baselineskip=24pt plus 2pt
\hfill{Revised Version}
\vspace{20mm}
\begin{center}
{\large \bf  Dyonic Black Holes in String Theory} \\
\vspace{30mm}
Guang-Jiun Cheng, Rue-Ron Hsu
\footnote{E-mail address: NCKUT057@TWNMOE10.EDU.TW.BITNET}
and Wei-Fu Lin
\vspace{20mm}
\\
Department of Physics,\\
National Cheng Kung University, \\
Tainan, Taiwan, 701, \\
Republic of China \\
\vspace{30mm}
\hfill {PACS:04.50.+h, 04.20.Jb}
\end{center}
\newpage

\begin{center}
{\bf ABSTRACT}
\end{center}
An exact solution of the low-energy string theory representing
static, spherical symmetric dyonic black hole is found.  The
solution is labeled by their mass, electric charge, magnetic
charge and asymptotic value of the scalar dilaton.  Some
interesting properties of the dyonic black holes are studied.
In particular, the Hawking temperature of dyonic black holes
depends on both the electric and magnetic charges, and the
extremal ones, which have nonzero electric and magnetic charges,
have zero temperature but nonzero entropy.  These properties are
quite different from those of electrically (or magnetically)
charged dilaton black holes found by Gibbons {\it et al.} and
Garfinkle {\it et al.}, but are the same as those of the
dyonic black holes found by Gibbons and Maeda.
\newpage
\section{Introduction}

Superstring theories are the most promising candidates for a
consistent quantum theory of gravity.  It is of interest to
investigate how the properties of black holes are modified when
the low-energy effective string actions are considered.  Recently,
some new black hole solutions have been obtained in the low-energy
string theories in which the Kalb-Ramond field, dilaton field and
gauge field are incorporated~\cite{1}-\cite{14}. Study of the
minimal coupling theory of gravity and the Kalb-Ramond field
indicate that there is no classical axionic hair in the
model~\cite{1}\cite{5}\cite{7}.  However, when dilaton and $U(1)$
gauge field are included, the new classical axionic hair have been
found~\cite{4}\cite{10}.  There are some charged black holes, with
electric or magnetic charge or dyon, in the effective string
theories in which gravity coupled to axion, dilaton field and
gauge field~\cite{4}\cite{10}-\cite{11}.
Reference~\cite{15}-\cite{17} are excellent reviews on the stringy
black holes.

Here, we consider the four-dimensional effective string action in
which gravity is coupled to dilaton and electromagnetic field only
\begin{equation}
   I = \int d^4x \sqrt{-g} \left[ - R + 2(\nabla \phi)^2
                                      + e^{-2\phi} F^2 \right]~~.
\end{equation}
Gibbons {\it et al.}~\cite{8} and Garfinkle {\it et al.}~\cite{9}
showed that the properties of electrically or magnetically charged
black holes can be drastically affected by incorporation of dilaton.
Gibbons and Maeda~\cite{8} also gave a dyonic black hole in the
effective string theory.  They reduced the field equations to a
`Toda molecule' form and obtained exact solutions by using a
peculiar radial coordinate.  Their solution is so complicated
that can not be understood by a easy way.  Therefore, it
is interesting to find an exact dyonic black hole
solution of the effective action, eq.(1), by using the usual radial
coordinate, and present it in a clear form.

In this paper, we find a static, spherical symmectric solution in
the low-energy effective string theory, eq.(1), in which gravity is
coupled to electromagnetic field and dilaton only.  It represents a
dyonic black hole which is characterized by their mass $M$, electric
charge $Q_e$, magnetic charge $Q_m$ and asymptotic value of the
dilaton $\phi_0$.  This solution is simpler and easier to interprete.
In analogy to the properties of Reissner-Nordstr\"om black hole, the
electrically (or magnetically) charged black holes found by Gibbons
and Maeda, and by Grafinkle, Horowitz and Strominger,
(GM-GHS solutions), are the limiting case of our results. Even the
electrically (or magnetically) charged black holes are the special
case of the dyonic black holes, however, unlike GM-GHS charged
black holes, the Hawking temperature of the dyonic black hole
depends on both the electric and magnetic charge, and vanishes
as $Q_e$ and $Q_m$ tend to extremal values.  These properties are
the same as those of the dyonic black holes found by Gibbons
and Maeda.

The plan of this paper is as follows. In Sec.2, we derive the
static, spherically symmectric Einstein field equations of the
theory eq.(1).  In Sec.3, we exhibit an exact dyonic black hole
solution.  We investigate the properties of the dyonic black holes
in Sec.4. Finally, we present some concluding remarks.

\section{The spherically symmetric field equations}

The field equations of the effective string action, eq.(1), are
\begin{equation}
   \nabla_{\mu} (e^{-2\phi} F^{\mu \nu})~~,
\end{equation}

\begin{equation}
   \nabla^2 \phi + \frac{1}{2} e^{-2\phi} F^2 = 0~~,
\end{equation}

\begin{equation}
   R_{\mu \nu} = 2\nabla_{\mu}\phi\nabla_{\nu}\phi
                 + 2e^{-2\phi}F_{\mu \rho} F_{\nu}^{~\rho}
                 - \frac{1}{2}g_{\mu \nu}e^{-2\phi}F^2~~.
\end{equation}
The most general spherical symmectric metric can be written in the
form \footnote{It is more convenient to generalize the electrically
(or magnetically) charged black holes to dyonic black holes by using
the spherically symmetric metric form, eq.(4), rather than that used
by Garfinkle {\it et al.}, $ds^2 = -\lambda(r^{\ast})^2 dt^2 +
\frac{1}{\lambda(r^{\ast})^2} {dr^{\ast}}^2 + R(r^{\ast})^2
\left( {d\theta^2} + {\sin}^2 \theta {d\varphi}^2 \right)~~.$}
\begin{equation}
   ds^2 = -\Delta^2 dt^2 + \frac{\sigma^2}{\Delta^2} dr^2
          + r^2 \left( {d\theta}^2
          + {\sin}^2 \theta {d\varphi}^2 \right)~~,
\end{equation}
where $t, r, \theta$ and $\varphi$ are ordinary spherical coordinates
and $\Delta$ and $\sigma$ are function of $r$ only.  Here, we
introduce the tetrad frame with basis 1-forms $\omega^0 = \Delta dt,
\omega^1 = \frac{\sigma}{\Delta} dr, \omega^2 = r d\theta$ and
$\omega^3 = r \sin \theta d\phi$.
The nonvanishing tetrad components of the $U(1)$ gauge field
strength~\cite{18}-\cite{20} are
\begin{equation}
   F_{01} = -F_{10} = f(r)~~,
\end{equation}
$$
   F_{23} = -F_{32} = g(r)~~,
$$
and the dilaton field $\phi$ depend on $r$ only.

Plugging eq.(5) and eq.(6) into eqs.(2)-(4), and the Bianchi
identity of gauge field
\begin{equation}
   \nabla_{\mu}F_{\nu \lambda} + \nabla_{\nu}F_{\lambda \mu} +
   \nabla_{\lambda}F_{\mu \nu} = 0~~,
\end{equation}
we end up with the following six equations
\begin{equation}
   (r^2 e^{-2\phi} f)' = 0~~,
\end{equation}

\begin{equation}
   (r^2 g)' = 0~~,
\end{equation}

\begin{equation}
   (\frac{r^2 \Delta^2}{\sigma} \phi')'
       = r^2 \sigma e^{-2\phi}(f^2-g^2)~~,
\end{equation}

\begin{equation}
   \left[ \frac{r \Delta}{\sigma} (r \Delta)'\right]' = \sigma~~,
\end{equation}

\begin{equation}
   (\frac{r \Delta^2}{\sigma} -\frac{r^2 \Delta\Delta'}{\sigma})'
   - \sigma = -2 r^2 \sigma e^{-2\phi} (f^2 + g^2)~~,
\end{equation}

\begin{equation}
   \left[ \frac{r \Delta}{\sigma}(r \Delta)'\right]'
   -2(\frac{2r\Delta}{\sigma} \Delta'+ \frac{\Delta^2}{\sigma})
   +\sigma = -2 r^2 \frac{\Delta^2}{\sigma} {\phi'}^2
   + 2 r^2 \sigma e^{-2\phi} (f^2 + g^2)~~,
\end{equation}
where the prime denotes differentiation with respect to $r$.

Eqs.(8) and (9) imply that
\begin{equation}
   f = \frac{Q_e}{r^2} e^{2\phi}~~,
\end{equation}
and
\begin{equation}
   g = \frac{Q_m}{r^2}~~,
\end{equation}
where $Q_e$ and $Q_m$ are integration constants and are relate to
electric and magnetic charge, respectively.

After replacing $f$,$~g$ by eqs.(14) and (15) and some suitable
recombinations of eqs.(10)-(13), we can reduce eqs.(10)-(13) to four
simpler equations
\begin{equation}
   (\frac{r^2 \Delta^2}{\sigma} \phi')' =
   \frac{\sigma}{r^2}({Q_e}^2 e^{2\phi} - {Q_m}^2 e^{-2\phi})~~,
\end{equation}

\begin{equation}
   \left[ \frac{1}{2}\frac{(r^2\Delta^2)'}{\sigma}\right]'=\sigma~~,
\end{equation}

\begin{equation}
   (\frac{r\Delta^2}{\sigma})'=\sigma \left[
   1-\frac{1}{r^2}({Q_e}^2 e^{2\phi} + {Q_m}^2 e^{-2\phi})\right]~~,
\end{equation}

\begin{equation}
   (\phi')^2 = \frac{1}{r}\frac{\sigma'}{\sigma}~~,
\end{equation}
which we will solve directly in the following section.

\section{Dyonic black hole solutions}

In this section, we will find the solutions of eqs.(16)-(19). The
key to solve these equations is eq.(19).  Consider the asymptotic
condition of $\phi$, $\phi (r \to \infty ) = \phi_0$, it implies
$\frac{\sigma'}{\sigma} \sim O(r^{-n})$, where $n > 1$, in
the asymptotic regime.  Therefore, we take a simply workable ansatz
\begin{equation}
   \frac{\sigma'}{\sigma} = \frac{\rho^2}{r(r^2 + \rho^2)}~~,
\end{equation}
where $\rho$ is a real constant.
Such that we get
\begin{equation}
   \phi' = \pm \frac{\rho}{r \sqrt{r^2 + \rho^2}}~~,
\end{equation}
where the different signs of $\phi'$ will give the same result
eventually, and we will study the ``$+$'' one only.
Solving eq.(17), (20) and (21), we can determine the functions
$\sigma (r)$, $\phi (r)$ and $\Delta (r)$ when the asymptotic
conditions, $\phi (r) \to \phi_0, \sigma (r) \to 1 $ and
$\Delta (r) \to 1$ as $r \to \infty$, are imposed.  eq.(16) gives us
some constraints on the integration constants in $\phi (r)$, $\Delta
(r)$ and $Q_e$, $Q_m$ and eq.(18) is automatically satisfied when
constraints on integration constants are imposed.

An exact solution of eqs.(16)-(19) is
\begin{equation}
   F_{01} = \frac{Q_e}{r^2} e^{2\phi}~~,
\end{equation}

\begin{equation}
   F_{23} = \frac{Q_m}{r^2}~~,
\end{equation}

\begin{equation}
   \sigma^2 = \frac{r^2}{r^2 + \rho^2}~~,
\end{equation}

\begin{equation}
   \Delta^2 = 1 - \frac{2 M}{r^2} \sqrt{r^2 + \rho^2}
                + \frac{\beta}{r^2}~~,
\end{equation}

\begin{equation}
   e^{2\phi} = e^{2\phi_0}(1-\frac{2 \rho}{\sqrt{r^2 + \rho^2}
               + \rho})~~.
\end{equation}
The constraints on the parameters are
\begin{equation}
   \rho = \frac{1}{2 M} ({Q_e}^2 e^{2\phi_0}
          - {Q_m}^2 e^{-2\phi_0})~~,
\end{equation}

\begin{equation}
   \beta = ({Q_e}^2 e^{2\phi_0} + {Q_m}^2 e^{-2\phi_0})~~.
\end{equation}
Here the integration constant $M$ is the mass of black holes, and is
determined by the asymptotical behavior of the metric.
The solutions of eqs.(22)-(28), with different parameters, are
related to one another by a simple dual transformation
\begin{equation}
   F \to \tilde{F} = \frac{1}{2} e^{-2\phi}
                      \epsilon_{\mu \nu}^{~~\lambda \rho}
                      F_{\lambda \rho}~~,
\end{equation}
$$
   \phi \to \tilde{\phi} = -\phi~~,
$$
which can be represented by $(Q_e,Q_m,\phi) \to (Q_m,Q_e,-\phi)$ in
our solution.

Moreover, we can find that the GM-GHS electrically (or magnetically)
charged black hole is a special case of our solution, eqs.(22)-(28),
in which $Q_m=0$ (or $Q_e=0$) but $Q_e \not= 0$ (or $Q_m \not= 0$).
For example, when we set $Q_e = 0$, the solution eqs.(22)-(28) will
reduce to magnetically charged black hole,
\begin{equation}
   F_{23} = \frac{Q_m}{r^2}~~,
\end{equation}

\begin{equation}
   \sigma^2 = \frac{r^2}{r^2 + {\tilde{\rho}}^2}~~,
\end{equation}

\begin{equation}
   \Delta^2 = 1 - \frac{2 M \sqrt{r^2 + \tilde{\rho}^2}}{r^2}
                + \frac{2 M \tilde{\rho}}{r^2}~~,
\end{equation}

\begin{equation}
   e^{-2\phi} = e^{-2\phi_0}(1-\frac{2 \tilde{\rho}}
               {\sqrt{r^2+{\tilde{\rho}}^2}+\tilde{\rho}})~~,
\end{equation}
where
\begin{equation}
   \tilde{\rho} = \frac{{Q_m}^2}{2 M} e^{-2\phi_0}~~.
\end{equation}
It is exactly the GM-GHS magnetic black hole solution
\begin{equation}
   F_{23} = \frac{Q_m}{{r^{\ast}}^2
            (1-\frac{2 \tilde{\rho}}{r^{\ast}})}~~,
\end{equation}

\begin{equation}
   e^{-2\phi} = e^{-2\phi_0}(1-\frac{2 \tilde{\rho}}{r^{\ast}})~~,
\end{equation}

\begin{equation}
   ds^2 = - (1-\frac{2 M}{r^{\ast}}) dt^2
          + \frac{1}{(1-\frac{2 M}{r^{\ast}})} {dr^{\ast}}^2
          + {r^{\ast}}^2 (1-\frac{2 \tilde{\rho}}{r^{\ast}})
          \left( d\theta^2 + \sin^2 \theta d\varphi^2 \right)~~,
\end{equation}
by a simple coordinate transformation
\begin{equation}
   r^{\ast} = \sqrt{r^2 + {\tilde{\rho}}^2} + \tilde{\rho}~~.
\end{equation}

\section{Properties of dyonic black hole solutions}

The structure of the dyonic black hole is similar to that of the
Reissner-Nordstr\"om one.  Three cases are considered as follows :

\begin{flushleft}
\begin{minipage}[t]{100mm}
{\bf CASE I}: DYONIC BLACK HOLE
\end{minipage}
\end{flushleft}

For $M > \frac{1}{\sqrt{2}} (|Q_e| e^{\phi_0} + |Q_m| e^{-\phi_0})$,
i.e., $2 M^2 - \beta > 2| Q_e || Q_m |$,
there are two zeros of $\Delta^2(r)$ at $r = r_{\pm}$, where
\begin{equation}
   r_{\pm} = {\left[ (2 M^2 - \beta) \pm  \sqrt{(2 M^2 - \beta)^2 -
   4 {Q_e}^2 {Q_m}^2}  \right] }^{1/2}~~,
\end{equation}
which correspond to two horizons.  The Kretschmann scalar
\begin{eqnarray*}
   K & = & R_{\mu \nu \lambda \tau} R^{\mu \nu \lambda \tau}\\
     & = & \frac{{\left[ (\Delta^2)'' \right]}^2}{\sigma^4}
     -(\Delta^2)'(\Delta^2)'' \frac{(\sigma^2)'}{\sigma^6}
     + ~\frac{1}{4} \frac{{\left[
     (\Delta^2)'(\sigma^2)' \right]}^2}{\sigma^8}
\end{eqnarray*}
$$
     + \frac{4}{r^2} \left\{ \frac{\left[\left(\Delta^2\right)'
     \right]^2} {\sigma^4} -\left(\Delta^2\right)'
     \left(\Delta^2\right) \frac{\left(\sigma^2\right)'}{\sigma^6}
     \right.
$$
\begin{equation}
     \left.
     +\frac{1}{2}\Delta^4\frac{\left[\left
     (\sigma^2\right)'\right]^2}{\sigma^8} \right\}
     +\frac{4}{r^4}\left(1-2\frac{\Delta^2}{\sigma^2}
     +\frac{\Delta^4}{\sigma^4} \right)~~,
\end{equation}
is finite at $r_{\pm}$ and is divergent at $r=0$, indicating
that $r_+$ and $r_-$ are regular horizons, and the singularity
locates at $r = 0$.  Since the Penrose diagram of these case is
the same as that of the Reissner-Nordstr\"om for $M > Q$, see
Fig.1(a).  Therefore, we may expect that $r_+$ coresponds to the
regular event horizon and $r_-$ coresponds to the unstable inner
horizon~\cite{21}.

The limiting case, $Q_e = 0$ or $Q_m = 0$, show that the inner
horizon will shrink to the singularity $r = 0$.  It means that the
electrically or magnetically charged black holes have only one
event horizon located at
\begin{equation}
   \tilde{r}_+ = \sqrt{4 M^2 - 2 {Q_e}^2 e^{2\phi_0}}~~,
\end{equation}
or
\begin{equation}
   \tilde{r}_+ = \sqrt{4 M^2 - 2 {Q_m}^2 e^{-2\phi_0}}~~,
\end{equation}
respectively, i.e., all of them are equivalent to
${r^{\ast}}_+ = 2 M$ in the coordinate used by Grafinkle
{\it et al.}.  The Penrose diagram of these limiting case is shown
in Fig.1(b).

\begin{flushleft}
\begin{minipage}[t]{100mm}
{\bf CASE II}: EXTREMAL DYONIC BLACK HOLE
\end{minipage}
\end{flushleft}

For $M = \frac{1}{\sqrt{2}} (|Q_e| e^{\phi_0} + |Q_m| e^{-\phi_0})$,
i.e., $2 M^2 - \beta = 2| Q_e || Q_m |$,
there is only one root of $\Delta^2(r) = 0$ at $r=r_0$
,where
\begin{equation}
   r_0 = \sqrt{2 M^2 -({Q_e}^2 e^{2\phi_0} + {Q_m}^2 e^{-2\phi_0})}=
         \sqrt{2 |Q_e||Q_m|}~~.
\end{equation}
Two horizons $r_+$ and $r_-$ match to form a regular event horizon.
$r = 0$ is still a singularity in this case.  The Penrose diagram of
the extremal dyonic black hole is shown in Fig.1(c).  According to
reference~\cite{22}, the extremal dyonic black hole solutions are
expected to be the end points of Hawking evaporation and coorespond
to stable quantum states.

We also see that $r_0 = 0$ if $Q_e = 0$ or $Q_m = 0$.  Since the
Kretschmann scalar diverges at $r_0 = 0$, the extremal electrically
or magnetically charged solution does not describe a black hole at
all but rather a naked singularity, see Fig.1(d).

\begin{flushleft}
\begin{minipage}[t]{100mm}
{\bf CASE III}: NAKED SINGULARITY
\end{minipage}
\end{flushleft}

For $M < \frac{1}{\sqrt{2}} (|Q_e| e^{\phi_0} + |Q_m| e^{-\phi_0})$,
the solution, eqs.(22)-(28), describles a naked singularity.
According to the cosmic censorship, this case should be forbidden,
and gives a extremal values for $Q_e$ and $Q_m$.
\vspace{10mm}

Besides investigate the structures of the black holes, we also study
the Hawking temperature of the dyonic black hole.
Based on Hawking's remarkable discovery - the laws of black hole
mechanics~\cite{23}, we find that the entropy $S$ and the Hawking
temperature $T_H$ of dyonic black holes are
\begin{equation}
   S = \frac{1}{4} ~(area) = \pi \left[ (2 M^2 - \beta) +
   \sqrt{(2 M^2 - \beta)^2 - 4 {Q_e}^2 {Q_m}^2} \right]~~,
\end{equation}
and
\begin{equation}
   T_H = \frac{1}{4 \pi M} \left[ \frac{\sqrt{(2 M^2 - \beta)^2 -
   4 {Q_e}^2{Q_m}^2}} {\sqrt{(2 M^2 - \beta)^2 -
   4 {Q_e}^2 {Q_m}^2}+(2 M^2 - \beta)}\right]~~.
\end{equation}
They depend on electric and magnetic charge of the black hole, and
show that the extremal dyonic black holes have non-zero entropy,
$S_0 = 2 \pi |Q_e||Q_m|$, at
zero temperature.  We may consider it as a result of a dual symmetric
which generates degenerate ground states of black hole.
Moreover, the electrically or magnetically charged black holes have
the charge independent Hawking temperature
$\tilde{T}_H = \frac{1}{8 \pi M}$ and the charge dependent entropy
$\tilde{S} = 4 \pi (M^2 - \frac{1}{2} {Q_e}^2 e^{2\phi_0})$ or
$\tilde{S} = 4 \pi (M^2 - \frac{1}{2} {Q_m}^2 e^{-2\phi_0})$.

The Hawking temperature of dyonic black holes is demonstrated
in Fig.2.  The temperature vanishes in the extreme limit on the
boundary except at the points $(\pm \sqrt{2} M e^{-\phi_0}, 0)$ or
$(0, \pm \sqrt{2} M e^{\phi_0})$.  The discontinuity of temperature
near the points $(\pm \sqrt{2} M e^{-\phi_0}, 0)$ or $(0,
\pm \sqrt{2} M e^{\phi_0})$, which correspond to extremal
electrically or magnetically charged solutions, reveals the
limitations on the thermal description of black holes~\cite{24}.
In Sec.5 we will see that the thermal description of the dyonic
black hole is inappropriate near the extreme limit.

\section{Concluding remarks}

In this paper, we give a dyonic black hole solution of the low-energy
effective string theory in which gravity is coupled to dilaton and
$U(1)$ gauge field.
These solutions are presented in an clear form by using usual
radial coordinate.
The structures and thermodynamic properties of dyonic black holes
are similar to those of the conventional charged black holes,
except for a purely electric or purely magnetic case.
Those properties are the same as those of dyonic black holes
found by Gibbons and Maeda.

As emphasized by Preskill {\it et al.}~\cite{24} and Holzhey
{\it et al.}~\cite{25}.  The description of a black hole as a thermal
object must break down as the extreme limit is approached.  They
suggest that the extreme pure electrically or pure magnetically
charged solutions should be regraded as elementary particles rather
than Reissner-Nordstr\"om black holes.  Here, we will check the self
consistent condition proposed by
Preskill {\it et al.} for the near-extreme dyonic black holes.

Due to
\begin{equation}
   {\left( \frac{\partial T}{\partial M} \right)}_{Q_e,Q_m} \simeq
   \frac{1}{2 \pi} \frac{1}{\sqrt{(2 M^2 - \beta)^2 -
   4 {Q_e}^2 {Q_m}^2}}~~,
\end{equation}
and the heat capacity of black hole,
\begin{equation}
   C_{Q_e,Q_m} \simeq 0
\end{equation}
as the extremal limit $2 M^2 - \beta = 2 |Q_e||Q_m|$ is approached.
The temperature fluctuation compared to the temperature itself,
\begin{equation}
\frac{\left< (\Delta T)^2 \right>}{T^2} = \frac{1}{C_{Q_e,Q_m}}
\end{equation}
will diverge wildly near the extreme regime.
It means that, under the assumption that the typical emitted quantum
radiation carries energy of order $T$ but no charge or angular
momentum, the self-consistent condition for the thermal description
\begin{equation}
   \biggl| T(\frac{\partial T}{\partial M})_{Q_e,Q_m} \biggr| \ll |T|
\end{equation}
is violated.  Therefore, we conclude that the thermodynamic
description of the near-extreme dyonic black hole is also
inappropriate.

\vspace{20mm}
\begin{center}
{\large \bf Note Added}
\end{center}
\vspace{10mm}
After this paper was submitted for publication, D. Wiltshire
told us that solutions, eqs.(22)-(28), are related to
Gibbons-Maeda dyonic black hole solutions by a coordinate
transformation and some parameters reparametization \cite{26}.
And, we were also informed that many of our results were previously
obtained by Kallosh {\it et al.} \cite{27}.
The dyonic black hole solutions, eqs.(22)-(28), are also
related to those of reference \cite{27} by another coordinate
transformation.

\vspace{20mm}
\begin{center}
{\large \bf Acknowledgements}
\end{center}
\vspace{10mm}
We thank Prof. D. Wiltshire who tells us some typographical errors
in \cite{8} and points out the relationship between our results and
the Gibbons-Maeda dyonic black hole solutions, and we also thank
Prof. A. Linde who tells us their works.
This research was supported in part by the National Science Council
of the Republic of China under grant No.~NSC 82-0208-M006-009
\newpage

\newpage

\begin{center}
{\large \bf Figure Captions}
\end{center}
\begin{flushleft}
\begin{minipage}[t]{13mm}
{\bf Fig.1}
\end{minipage}
\begin{minipage}[t]{120mm}
The Penrose diagrams of dyonic black hole with a dilaton
and its limiting cases, (a) dyonic black hole with both electric and
magnetic charge, (b) electrically or magnetically charged black hole,
(c) extremal dyonic black hole, (d) extremal electrically or
magnetically charged solution does not describe a black hole at all
but rather a naked singularity.
\end{minipage}\\
\vspace{10mm}
\begin{minipage}[t]{13mm}
{\bf Fig. 2}
\end{minipage}
\begin{minipage}[t]{120mm}
The Hawking temperature $T_H$ of dyonic black holes v.s. $Q_e$ and
$Q_m$.  Inside the square, we find a regular dyonic black hole.
On the boundary except for points
$\left( \pm \sqrt{2} M e^{-\phi_0}, 0 \right)$ or
$\left(0, \pm \sqrt{2} M e^{\phi_0} \right)$ or, we
find an extremal dyonic black hole.  On the $Q_e$ or $Q_m$
axis, we find the electrically or magnetically charged black hole.
The temperature is finite on the axes.
Finally, points
$\left( \pm \sqrt{2} M e^{-\phi_0}, 0 \right)$ or
$\left(0, \pm \sqrt{2} M e^{\phi_0} \right)$
represent the extremal electrically or
magnetically charged solutions.
\end{minipage}
\end{flushleft}


\begin{thebibliography}{set}

\bibitem{1}
M.J. Bowick, S.B. Giddings, J.A. Harvey, G.T. Horowitz and
A. Strominger, {\bf Phys. Rev. Lett.~} {\bf 61} (1988) 2823.
\bibitem{2}
M.J. Bowick, {\bf Gen. Rel. Grav.~} {\bf 22} (1990) 137.
\bibitem{3}
B.A. Campbell, M.J. Duncan, N. Kaloper and K.A. Olive,
{\bf Phys. Lett.~} {\bf B251} (1990) 34,
{\bf Nucl. Phys.~} {\bf B351} (1991) 778.
\bibitem{4}
B.A. Campbell, N. Kaloper and K.A. Olive, {\bf Phys. Lett.~}
{\bf B263} (1991) 364, {\bf B285} (1992) 199.
\bibitem{5}
R.R. Hsu, {\bf Class. Quantum Grav.~} {\bf 8} (1991) 779.
\bibitem{6}
R.R. Hsu and W.F. Lin, {\bf Class. Quantum Grav.~} {\bf 8} (1991)
L161.
\bibitem{7}
R.R. Hsu, G. Huang and W.F. Lin, ``{Is there classical axionic hair
in the minimal coupling theory of gravity and Kalb-Ramond field,}''
preprints, to appear in {\bf Chinese J. Phys.} (Taipei).
\bibitem{8}
G.W. Gibbons and K. Maeda, {\bf Nucl. Phys.~} {\bf B298} (1988) 741.
\bibitem{9}
D. Garfinkle, G.T. Horowitz and A. Strominger, {\bf Phys. Rev.~}
{\bf D43} (1991) 3140.
\bibitem{10}
A. Shapere, S. Trived and F. Wilczek, {\bf Mod. Phys. Lett.~}
{\bf A6} (1991) 2677.
\bibitem{11}
K. Lee and E. Weinberg, {\bf Phys. Rev.~} {\bf D44} (1991) 3159.
\bibitem{12}
T. Koikawa and M. Yoshimura, {\bf Phys. Lett.~} {\bf B189} (1987) 29.
\bibitem{13}
I. Ichinose and H. Yamazaki, {\bf Mod. Phys. Lett.~} {\bf A4}
(1991) 1509; H. Yamazaki and I. Ichinose, {\bf Class. Quantum Grav.~}
{\bf 9} (1992) 257.
\bibitem{14}
J.H. Horne and G.T. Horowitz, {\bf Phys. Rev.~} {\bf D46}
(1992) 1340.
\bibitem{15}
G.T. Horowitz, ``{The Dark Side of String Theory: Black Holes and
Black Strings,}'' UCSBTH-92-32 (1992).
\bibitem{16}
G.T. Horowitz, ``{What is the True Description of Charged Black
Hole,}'' UCSBTH-92-52 (1992).
\bibitem{17}
J.A. Harvey and J.A. Strominger, ``{Quantum Aspects of Black
Holes,}'' EFI-92-41 (1992).
\bibitem{18}
S. Weinberg, ``{\it Gravitation and Cosmology,}''
New York, Wiley (1972) Ch.13.
\bibitem{19}
R. Rauch and H.T. Nieh, {\bf Phys. Rev.~} {\bf D24} (1981) 2029,
Appendix B.
\bibitem{20}
S. Chandrasekhar, ``{\it The Mathematical Theory of Black Hole,}''
Oxford University Press (1983).
\bibitem{21}
J. McNamara, {\bf Proc. Roy. Soc. Lond.~} {\bf A358} (1978) 449;
Y. Gursel, V. Sandberg, I. Novikov and A. Starobinski,
{\bf Phys. Rev.~} {\bf D19} (1979) 413; R. Matzner, N. Zamorano
and V. Sandberg, {\bf Phys. Rev.~} {\bf D19} (1979) 2821;
S. Chandrasekhar and J. Hartle, {\bf Proc. Roy. Soc. Lond.~}
{\bf A384} (1982) 301.
\bibitem{22}
S.W. Hawking, {\bf Commun. Math. Phys.~} {\bf 43} (1975) 199.
\bibitem{23}
S.W. Hawking, {\bf Phys. Rev.~} {\bf D13} (1976) 191.
\bibitem{24}
J. Preskill, P. Schwarz, A. Shapere, S. Trivedi and F. Wilczek,
{\bf Mod. Phys. Lett.~} {\bf A6} (1991) 2351.
\bibitem{25}
C. Holzhey and F. Wilczek, {\bf Nucl. Phys.~} {\bf B380} (1992) 447.
\bibitem{26}
D. Wiltshire, private communication.
\bibitem{27}
R. Kallosh, A. Linde, T. Ortin, A. Peet and A. Van Proeyen,
{\bf Phys. Rev.~} {\bf D46} (1992) 5278.

\end{thebibliography}
\end{document}